\documentclass[conference]{IEEEtran}

\newtheorem{theorem}{\bf Theorem}

\usepackage{amsmath} 
\usepackage{algorithm}
\usepackage{algpseudocode}
\usepackage{pifont}
\usepackage{graphicx}
\usepackage{afterpage}
\usepackage{float}
\graphicspath{{images/}}
\usepackage{subcaption}
\usepackage{xcolor}
\usepackage{gensymb}
\usepackage{qtree} 

\ifCLASSINFOpdf
\else
\fi
\begin{document}
%
\title{Non-linear Barrier Coverage using Mobile Wireless Sensors}

\author{\IEEEauthorblockN{Ashutosh Baheti}
\IEEEauthorblockA{Dept. of Computer Sc. \& Engineering\\
Indian Institute of Technology, Kharagpur\\
Kharagpur, WB-721302, India\\
Email: ashutosh.baheti95@gmail.com}
\and
\IEEEauthorblockN{Arobinda Gupta}
\IEEEauthorblockA{Dept. of Computer Sc. \& Engineering\\
Indian Institute of Technology, Kharagpur\\
Kharagpur, WB-721302, India\\
Email: agupta@cse.iitkgp.ernet.in}}


%


\maketitle
\thispagestyle{plain}

\begin{abstract}
A belt region is said to be \emph{k-barrier covered} by a set of sensors if all paths crossing the width of the belt region intersect the sensing regions of at least $k$ sensors. Barrier coverage can be achieved from a random initial deployment of mobile sensors by suitably relocating the sensors to form a barrier. Reducing the movement of the sensors is important in such scenarios due to the energy constraints of sensor devices. In this paper, we propose a centralized algorithm which achieves 1-barrier coverage by forming a non-linear barrier from a random initial deployment of sensors in a belt. The algorithm uses a novel idea of physical behavior of chains along with the concept of virtual force. Formation of non-linear barrier reduces the movement of the sensors needed as compared to linear barriers. Detailed simulation results are presented to show that the proposed algorithm achieves barrier coverage with less movement of sensors compared to other existing algorithms in the literature.
\end{abstract}


%
\IEEEpeerreviewmaketitle

\section{Introduction}
 A Wireless Sensor Network (WSN) consists of a set of sensor nodes. Each node can sense some physical parameters, and has limited computation and communication capability. The data sensed by the sensor nodes is usually transmitted to a central base station using a wireless network connecting the sensors for further processing. Mobile Wireless Sensor Networks (MWSNs) \cite{wang2006movement}, \cite{ma2008managing}, \cite{bartolini2011autonomous} are a class of wireless sensor networks in which some or all of the sensor nodes are mobile. Wireless sensor networks have been used for a wide variety of applications such as habitat monitoring, target tracking, intruder detection etc.
\par 
 Many different problems have been addressed on wireless sensor networks, such as routing, topology control, localization, coverage, data aggregation etc. In this work, we focus on the coverage problem in sensor networks, which addresses the problem of covering a given area or set of objects using sensors. Several different variations of the problem exist depending on the nature of coverage required. As an example, \textit{Area Coverage} requires that all points in a given area are within the sensing field of at least one sensor. In contrast, \emph{Point Coverage} requires that only a given set of points within an area be covered by at least one sensor. Many other definitions of coverage exist such as perimeter coverage, barrier coverage, sweep coverage, path coverage etc. \cite{Wang2011}.

\par 
In this paper, we focus on  \textit{Barrier Coverage} in sensor networks.
A target belt region provides \emph{strong k-barrier coverage} if all \emph{crossing paths} (a path crossing the width of the belt region, originating from one parallel boundary of the belt region and terminating at the other) intersect the sensing region of at least \emph{k} distinct sensors. In the rest of this paper, we will refer to strong barrier coverage as just barrier coverage. Barrier coverage has many important applications such as intrusion detection along international borders, identifying spread of lethal chemicals around chemical factories, detecting potential sabotage in gas pipelines etc. \cite{chen2010local}, \cite{saipulla2013barrier}.

\par 
Static WSNs may not always work well in barrier coverage applications for different reasons. In many applications, sensors cannot be placed exactly at the locations desired due to deployment constraints, and hence sensors may be randomly deployed around the area. Such applications can benefit from mobile WSNs if the deployed sensors can autonomously move after deployment to achieve the desired barrier coverage. Mobile sensors can also help in scenarios where one or more sensors fail, thereby breaking barrier coverage. In such scenarios, some of the nearby sensors can readjust their positions to recreate the barrier. However, designing algorithms that utilize mobility efficiently is a challenging problem. Keeping the energy constraint of battery-powered sensor devices in mind, algorithms for mobility control should be able to achieve barrier coverage with low movement of the sensors. 
\par 
There exist many works on formation of barrier coverage using mobile sensors. Some of these works propose centralized solutions for deploying sensors to achieve barrier coverage  \cite{bhattacharya2009optimal}, \cite{saipulla2010barrier}, \cite{saipulla2013barrier}. Centralized solutions are good since all the information is available at a central station. Hence, all computations regarding movement can be done centrally and only the final locations are sent to the sensors which then move to those locations. However, maintaining the information centrally incurs some overhead. To address this problem, distributed approaches in which sensors locally coordinate to adjust and move to their final positions have also been proposed \cite{shen2008barrier}, \cite{cheng2009problem}, \cite{yang2007defending}, \cite{kong2010automatic}, \cite{silvestri2011mobibar}. However, most of these solutions (centralized or distributed) try to organize all sensors in a straight line to achieve barrier coverage. While  a linear barrier is optimal with respect to the number of sensors needed to create a barrier, it can cause large redundant movement to move randomly deployed sensors to a linear configuration.
\par 
In this paper, we propose a centralized algorithm for 1-barrier coverage which, given a random deployment of sensors in a belt, creates a non-linear barrier of sensors that provides barrier coverage of the belt region. The proposed algorithm views the barrier as a chain of sensors whose sensing disks overlap with each other, and uses some novel ideas of physical behavior of a chain along with the concept of virtual force to move the sensors to achieve barrier coverage.  Detailed simulation results are presented to show that the proposed algorithm achieves barrier coverage from random initial deployments with both less average displacement  and less maximum displacement of the sensors compared to other existing algorithms in the literature.
\par 
The rest of the paper is organized as follows. Section \ref{related} discusses some related works. The problem statement is defined in Section \ref{problem}, Section \ref{centalgo} describes the centralized algorithm and evaluates its performance. Finally, Section \ref{conclusion} concludes the work.

\section{Related Works}
\label{related}
Barrier coverage has been widely studied in WSNs \cite{cardei2006energy}, \cite{fan2010coverage}. Saipulla et al. \cite{saipulla2013barrier} suggested an approach to relocate sensors from an initial randomized line-based deployment model to replicate a scenario of sensors dropped from an aircraft. The work in \cite{saipulla2010barrier} studied the problem of relocating mobile sensors with limited mobility in order to save energy. Bhattacharya et al. \cite{bhattacharya2009optimal} proposed and solved an optimization problem to calculate an optimal movement strategy for barrier coverage on a circular region. All these approaches are centralized. Distributed approaches to barrier coverage formation are addressed in \cite{kong2010automatic}, \cite{silvestri2011mobibar}, \cite{shen2008barrier}, \cite{cheng2009problem}, \cite{Eftekhari2013}. Kong et al. \cite{kong2010automatic} used the concept of virtual forces to solve the \emph{k-barrier coverage}. Silvestri \cite{silvestri2011mobibar} presented a novel approach \emph{MobiBar} that outperforms the algorithm of  \cite{kong2010automatic} in $k$-barrier coverage. Cheng and Savkin \cite{cheng2009problem} presented a decentralized approach of creating 1-barrier coverage between any two pre-specified landmarks in the belt region. Shen et al. \cite{shen2008barrier} suggested a centralized \emph{CBarrier} and a distributed \emph{DBarrier}(based on virtual forces) algorithm that create a barrier after an random initial deployment. Eftekhari et al. \cite{Eftekhari2013} presented distributed algorithms for barrier coverage using sensor relocation.  All of the algorithms proposed in the literature (except \emph{DBarrier} whose performance has been shown to be poorer than \emph{CBarrier}) try to rearrange sensors on a straight line to form a linear barrier. Ban et al. \cite{Ban2010} defined a special type of non-linear $k$-barrier coverage called \textit{grid barrier} that is formed out of linear segments coinciding with grid boundaries in the region; however, the algorithm presented for 1-barrier coverage only (named \textit{CBGB}) still forms linear barrier only. A linear barrier uses the minimum possible number of sensors, but causes more movement of the sensors to arrange them along a straight line. When some extra sensors (over the minimum number required) are available, forming a non-linear barrier can reduce the movement of the sensors and consequently cause less energy usage. Ban et al. \cite{Ban2010} also presented a more general $k$-barrier coverage algorithm that creates a non-linear grid barrier by breaking the region into subregions, forming linear barriers in each subregions, and then forming vertical isolation barriers between subregions to connect the horizontal barriers. However, the algorithm uses very large number of redundant sensors, and for 1-barrier coverage ($k=1$), provides no significant advantage over the \textit{CBGB} algorithm.
The algorithm proposed in this paper differs from the other algorithms (except \emph{DBarrier} \cite{shen2008barrier}) in that it finds truly non-linear barriers which reduces the movement of the sensors. The number of redundant sensors used is also low.

\section{Problem Formulation}
\label{problem}
We assume a rectangular belt region of length $L$ and width $W$,  with $L \gg W$. A set of $N$ mobile sensors with unique IDs are initially randomly deployed in this belt region. Each sensor has a sensing range $R_s$ and thus can cover a circular area of radius $R_s$ centered around the position where the sensor is placed. 
We assume that a sensor knows its own location. 
\par
The displacement of a sensor is defined as the Euclidean distance between the initial and final location of the sensor. As noted earlier, a centralized algorithm can compute the final location of the sensors and then the sensors can move to that location directly. Thus, the goal of the centralized barrier formation algorithm is to relocate the sensors to form 1-barrier coverage over the belt region while minimizing the average and maximum displacement of the sensors. 

\section{Centralized Algorithm for Barrier Formation}
\label{centalgo}
We first describe the intuition behind the proposed algorithm. A more formal description of the algorithm is given next.
\par
A barrier can be viewed as a physical chain where each sensor, with its imaginary sensing disk, is analogous to a circular chain link. In a physical chain, if one chain link is pulled, it will exert a force on all chain links connected to it, and a chain link connected to it will move when the distance between them becomes maximum, i.e., when their rims start touching. We primarily use this simple property in designing the algorithm. Some of the terms we will use to describe the operation of the algorithm are described below. 
\begin{itemize}
    \item \textbf{Chain Link}: A \textit{chain link} in the algorithm refers to a sensor with sensing radius $R_s$. We will use the terms sensor and chain link interchangeably in the rest of this paper. The leftmost sensor in the belt region is called the \textit{left chain link}. Similarly, the rightmost sensor is called the \textit{right chain link}. In case there are multiple leftmost (rightmost) sensors, any one sensor is chosen as the left (right) chain link.
    \item \textbf{Connected Chain Links}: Two chain links are said to be \textit{connected} if the sensing regions of the corresponding sensors intersect. Note that the distance between the centers of two connected chain links cannot be more than $2R_s$. If one chain link is pulled, a chain link connected to it will feel a force when the distance between their centers becomes  equal to $2R_s$.
    \item \textbf{Chain Graph}: Consider the undirected graph $G$ with each chain link as a node and an edge added between two nodes if the corresponding chain links are connected. A chain graph is a connected component of $G$. The chain graph that contains the left chain link is called the \textit{left chain graph}. Similarly, the chain graph containing the right chain link is called the \textit{right chain graph}.
\end{itemize}

Note that though a distance less than $2R_s$ between the positions of two sensors implies that the sensors are connected by the above definitions, sometimes we will \emph{delete} a connection even if the distance is less than $2R_s$ to work with a subgraph of the chain graph. Such deletions will be clearly specified while describing the algorithm.

 Given the above virtual constructs, the main idea of the algorithm can be described as follows. Each chain graph, by virtue of its connectedness, can provide barrier coverage for a part of the belt region (the part between the left boundary of its leftmost chain link and the right boundary of its rightmost chain link). The algorithm tries to extend the barrier coverage to the entire belt region by merging smaller chain graphs into larger ones, until a large enough chain graph is formed that spans across the length of the belt. In order to merge two chain graphs, one special chain link of one chain graph is pulled towards a chain link of the other chain graph in steps, pulling the other chain links along with it until the two chains merge. We next describe this technique in more detail.

\subsection{Merging Chain Graphs}
Let $C$ denote the set of all chain links (sensors). Let $C_G$ denote the set of chain links in a chain graph $G$. For any chain link $u \in C_G$ and $v \in C \setminus C_G$, we define the force $f(u,v)$ exerted on $u$ by $v$ as  
\begin{align*}
f(u,v) &= \frac{\alpha}{dist(u,v)} & u \in C_G, \;\; v \in C\setminus C_G
\end{align*}
where $dist(u,v)$ is the Euclidean distance between $u$ and $v$. Note that $f$ represents the attractive force between two chain links in two chain graphs, with the force becoming stronger as the distance between the chain links decreases. $\alpha$ is simply a scaling parameter.

For a chain link $u$ in $C_G$, let 
\begin{align*}
F(u) &= \underset{v}{\text{ max }} f(u,v) & u \in C_G ,\;\;\forall~ v \in C\setminus C_G
\end{align*}
denote the maximum force exerted on $u$ by a chain link in another chain graph. Then the \textit{dominant point} of the chain graph $G$ is defined as the chain link $d_G \in C_G$ such that $F(d_G)$ is the maximum among all $F(u), u \in C_G$. The chain link in a chain graph $X \neq G$ that exerts the maximum force on $d_G$ is defined as the \textit{co-dominant point} of $G$. If there are more than one pair of dominant and co-dominant points with the maximum force then we choose the pair which has a sensor with the lowest ID.

\par
 For merging two chain graphs, we compute the dominant and the co-dominant points of the chain graphs and then pull the dominant point towards the co-dominant point using the attractive force defined, pulling the other chain links in the chain graph along with it. The movements of a chain link due to the attractive force as well as when a connected chain link is pulled follows the usual laws of physics. The algorithm for merging chain graphs works in four phases:

\begin{enumerate}
    \item \textbf{Initialization Phase:}
     In this phase, the algorithm first identifies all the chain graphs based on the initial locations of the sensors. For each chain graph, a DFS spanning tree is constructed starting from an arbitrary node, and all edges not in the tree are deleted from the chain graph. Removing an edge between two chain links implies that they are not considered as connected (even if their sensing regions overlap) in the rest of the algorithm and are not constrained to move together in any way. The DFS tree helps in identifying long paths in chain graphs that will be used to flatten the chain graphs later. At the end of this stage, all chain graphs to be considered are trees. We will refer to the DFS tree corresponding to a chain graph as a \textit{chain tree} in the rest of this section.
    \item \textbf{Left-Attach Phase:}
    In this phase, the left chain link is pulled horizontally towards the left boundary of the belt until it touches the left boundary. Note that pulling the left chain link may pull other chain links in the left chain graph. Once the left chain link touches the left boundary, it is allowed to only slide along the width of the belt but is not allowed to move away from the left boundary in the rest of the algorithm.  
    \item \textbf{Right-Attach Phase:}
    This phase is similar to the Left-Attach Phase, only difference being the right chain link is moved horizontally towards the right boundary of the belt in this case.
    \item \textbf{Barrier-Formation Phase:}
    This phase actually creates the barrier by merging chain graphs. The dominant point and co-dominant points are first determined for all chain graphs. The algorithm then iterates over the following step, with every iterative step corresponding logically to a time step $\tau$ of movement defined suitably. 
	\begin{enumerate}
	\item
	One chain graph $G$ is picked up randomly from the set of chain graphs.
	\item
	The dominant point $d_G$ of $G$ is moved towards the co-dominant point of $G$ for duration $\tau$ (or until it touches the co-dominant point, whichever is earlier) following the laws of physics depending on the force $f$ between them. Note that this may move other chain links in $G$, again following the laws of Physics.
	\item
	If $d_G$ does not touch the co-dominant point after the movement, the force $f$ between $d_G$ and the co-dominant point is recomputed (as it may have changed due to the change in distance between them), and the algorithm goes to the next iteration. If $d_G$ touches the co-dominant point (the two chain graphs have merged), then the dominant and co-dominant point of all the chain graphs are recomputed before going to the next iteration.
	\end{enumerate}
This phase terminates when the merging of the chain graphs causes a barrier to be formed. Thus, at each step of this phase, one chain graph is moved slightly. Note that moving one arbitrary chain graph completely in one step to cause two chain graphs to merge may cause a lot of redundant movement unless the proper chain graph is chosen; the random choice reduces this redundant movement in case of a bad choice of chain graph.
	
\end{enumerate}

Note that a chain graph can potentially cover a larger part of the belt if more nodes in it have degree two, as higher degree nodes in a chain graph cause more redundant nodes that do not contribute in extending the barrier. Therefore, at each step of moving the dominant point, a \textit{flattening logic} is applied to the chain graph that flattens the chain to make it longer. 

\subsection{Flattening Chain Graphs}
 The flattening logic is applied to the chain tree formed from the chain graph.  
For any chain tree, the dominant point of the chain tree is defined as the \textit{root chain link} of the tree. For any chain tree (except the left and the right chain trees) the longest path from the root chain link in the chain tree is first computed. This path is called the \textit{flatten path}. The following step are applied after each step of the Barrier Formation phase. 

The flatten path is traversed starting from the root chain link until the first node with degree greater than two (branching) is found. Let this node be called the \textit{current chain link} $c$. Let $N_c$ denote the set of chain links connected to $c$ and $S_c$ denote the set of chain links connected to $c$ but are not on the flatten path. For every chain link $u \in S_c$, the chain link $v$ ($v \in N_c \setminus S_c$) closest to $u$ in the flatten path is found, and a small fixed attractive force (taken to be $\beta \times f$, $0 < \beta << 1$, where $f$ is the force between the dominant and the co-dominant point of the chain graph) is exerted upon $u$ towards $v$ to make $u$ move towards $v$. If $u$ touches $v$ as a result of this movement, an edge is added in the chain graph between $u$ and $v$ and the edge between $c$ and $v$ is deleted (even if their sensing regions overlap). Note that this extends the flatten path by one sensor (replacing the edge $(c,v)$ in the flatten path with the edge $(c,u)$ and $(u,v)$). The change in edges still maintains a tree. 

Thus, as the dominant point of a chain graph moves towards another chain graph in successive steps, the application of the above logic brings more and more nodes into the flatten path, eventually causing all chain links not present in this flatten path to  collapse on it. This, along with the movement of the dominant point which pulls the chain links on the flatten path, causes a chain to become longer with sufficient number of steps, thus allowing a single chain to form a larger part of the barrier. Note that for a long enough belt region (large $L$), if the merging step and the flattening logic is applied on a chain graph long enough, the graph will be transformed into a linear chain with all nodes having degree two (except the end nodes that are with degree one),  with the distance between the centers of two connected chain links becoming maximum.

\begin{figure}
    \begin{subfigure}{.49\linewidth}
        \centering
        \includegraphics[width=\linewidth]{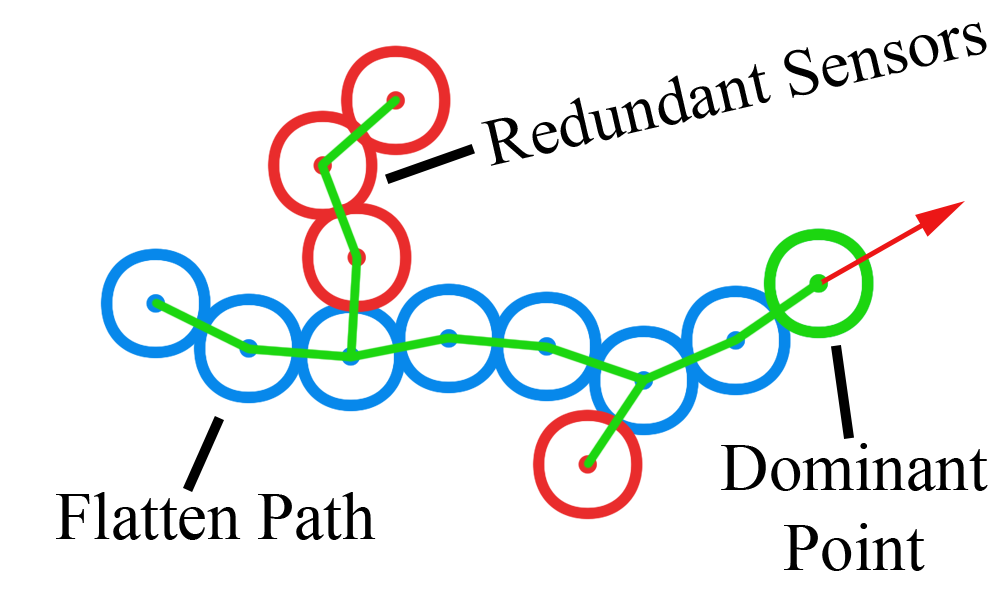}
        \caption{}
        \label{fig:sfig1}
    \end{subfigure}
    \begin{subfigure}{.49\linewidth}
        \centering
        \includegraphics[width=\linewidth]{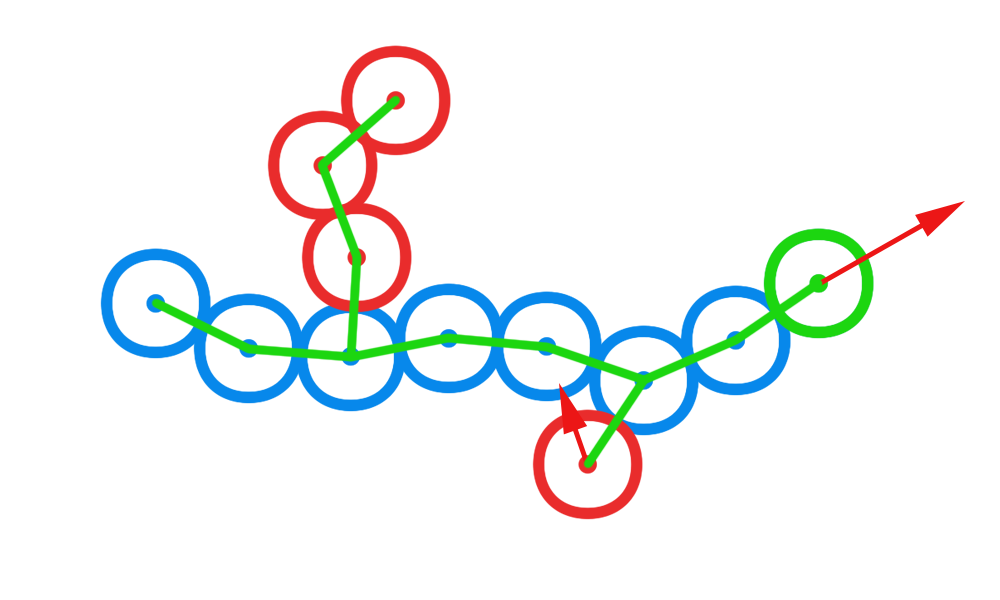}
        \caption{}
        \label{fig:sfig2}
    \end{subfigure}\\
    \begin{subfigure}{.49\linewidth}
        \centering
        \includegraphics[width=\linewidth]{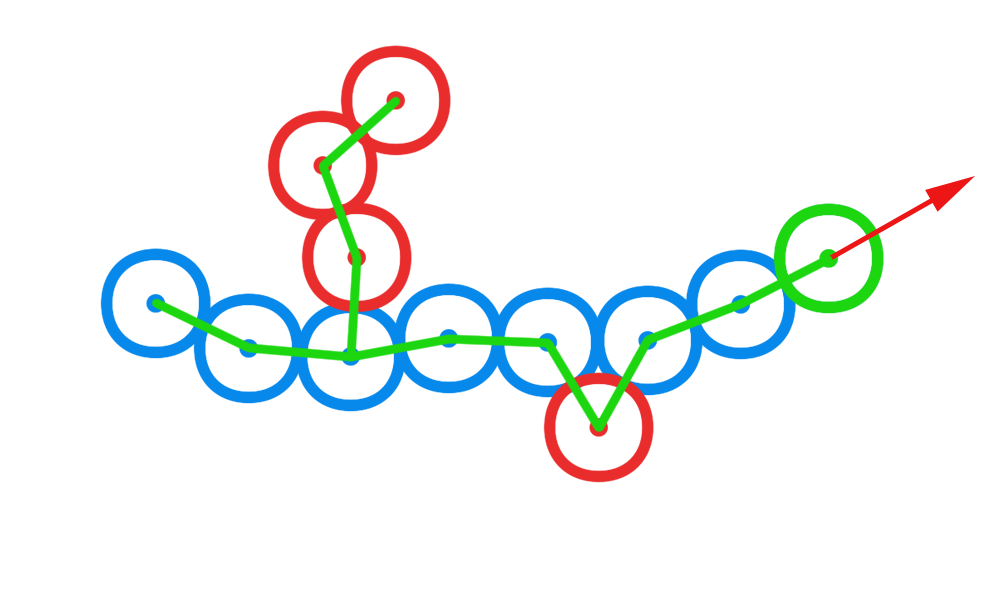}
        \caption{}
        \label{fig:sfig3}
    \end{subfigure}
    \begin{subfigure}{.49\linewidth}
        \centering
        \includegraphics[width=\linewidth]{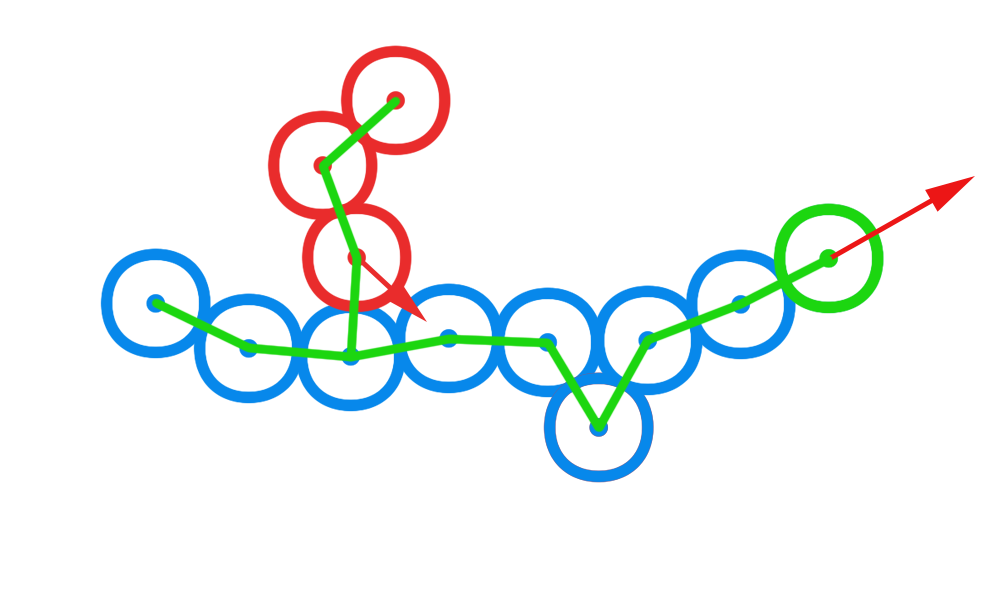}
        \caption{}
        \label{fig:sfig4}
    \end{subfigure}\\
    \begin{subfigure}{.49\linewidth}
        \centering
        \includegraphics[width=\linewidth]{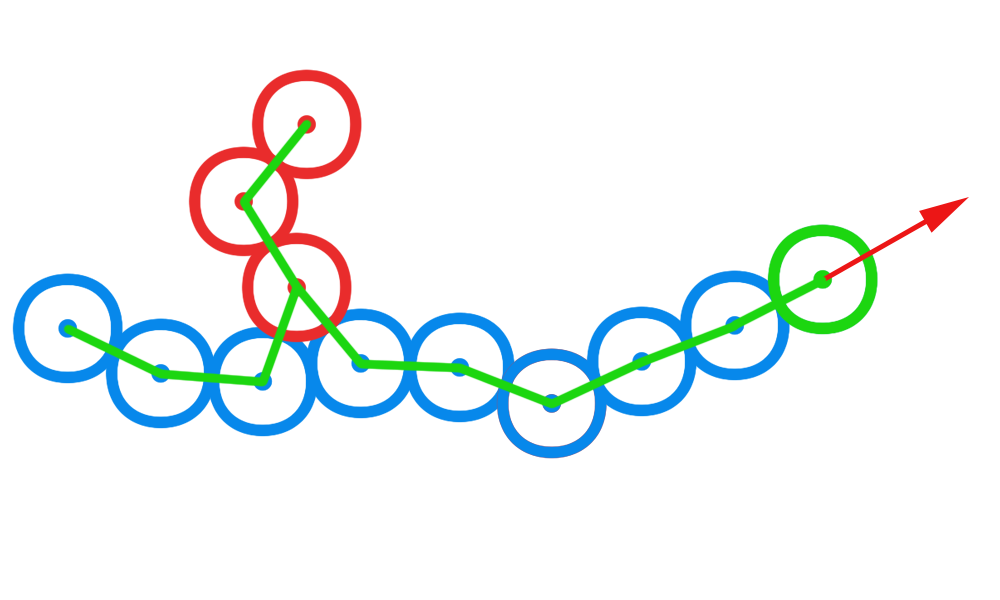}
        \caption{}
        \label{fig:sfig5}
    \end{subfigure}
    \begin{subfigure}{.49\linewidth}
        \centering
        \includegraphics[width=\linewidth]{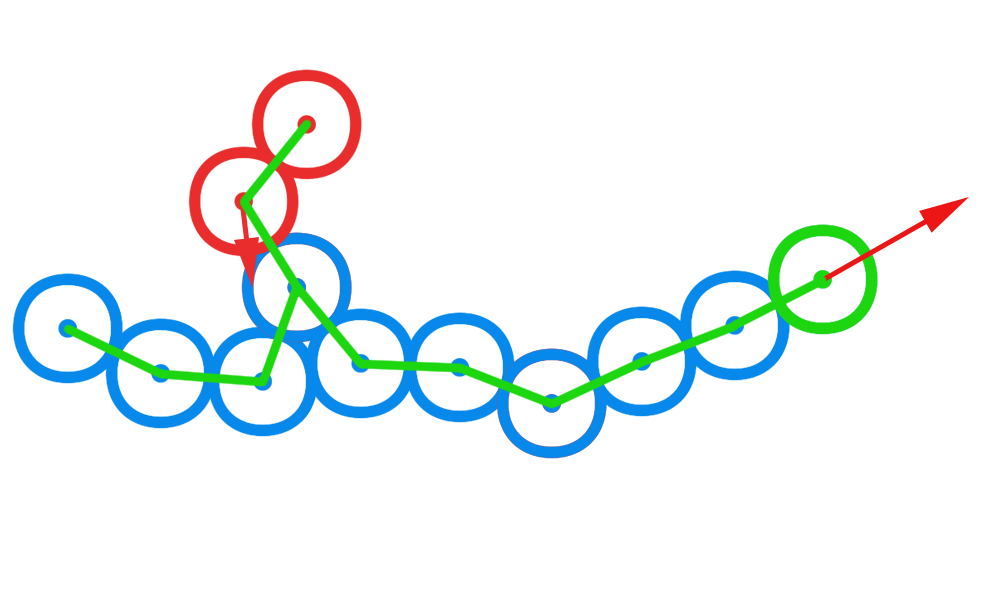}
        \caption{}
        \label{fig:sfig6}
    \end{subfigure}\\
    \caption{Example of the flattening logic}
    \label{fig:flattening}
\end{figure}
\par
 An example of the flattening logic is shown in Figure \ref{fig:flattening}. Figure \ref{fig:sfig1} shows the initial state of a chain tree. The green chain link denotes the dominant point, with the red arrow depicting the direction of the force exerted. The chain links on the flatten path computed from the root chain link (dominant point) are shown in blue. As shown in Figure \ref{fig:sfig2}, a node connected to the the first node on the flatten path (starting from the root) with degree greater than two is moved towards the flatten path. When this node meets the flatten path after sufficient number of steps, the flatten path is changed as shown in  Figure \ref{fig:sfig3}. This adds one more sensor to the flatten path. At later steps, the next chain link on the flatten path with degree greater than two is chosen and a chain link connected to it is moved towards and finally added to the flatten path (Figures \ref{fig:sfig4} and \ref{fig:sfig5}). The next chain link on this new flatten path with degree greater than two is then moved towards the flatten path (Figure \ref{fig:sfig6}). This process is repeated with each step to flatten the chain tree as much as needed. Note that as the dominant point moves because of the force exerted on it, the chain links on the flatten path are pulled along with it, causing the chain to straighten out and form a longer barrier.
\par
 For the left and the right chain tree, the flatten path is fixed to be the path from the dominant point to the left and the right chain links respectively. The rest of the logic remains the same. This is done because of the special constraint on the leftmost and the rightmost sensors. The pseudo code for the flattening logic is given in Algorithm \ref{FlatteningAlgo}. More details are omitted because of space constraints.

\begin{algorithm}
\caption{Flattening Algorithm}
\label{FlatteningAlgo}
    \begin{algorithmic}[1]
    \Function{flatten}{chain graph $G$, chain link $root$}
        \State Compute flatten path $P$ from $root$
        \State \textit{current} = $root$
        \While{ degree of $current$ $\le$ 2}
            \State \textit{current} = next node after \textit{current} on $P$    \EndWhile
        \State $S_c$ = set of neighbor nodes of $current$ that are not on $P$
        \For{each $u \in S_c$}
            \State $v$ = neighbor node of \textit{current} closest to $u$ on $P$
            \State apply force $f$ on $u$ towards $v$ to move $u$
                \If{distance between $u$ and $v$ $\le$ $2R_s$}
                    \State delete edge $(current,v)$ from $G$
                    \State add edge $(u,v)$ to $G$
        		    \State $P = P \cup \{(u,v), (current,u)\} - {(current, v)}$
        		    \State break
                \EndIf
        \EndFor
        
    \EndFunction
    \end{algorithmic}
\end{algorithm}

\begin{figure}
    \begin{subfigure}{\linewidth}
        \centering
        \includegraphics[width=.9\linewidth]{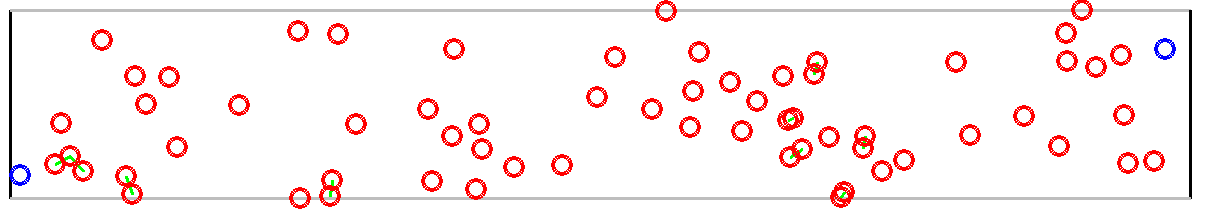}
        \caption{}
        \label{fig:efig1}
    \end{subfigure}
    \par \medskip
    \begin{subfigure}{\linewidth}
        \centering
        \includegraphics[width=.9\linewidth]{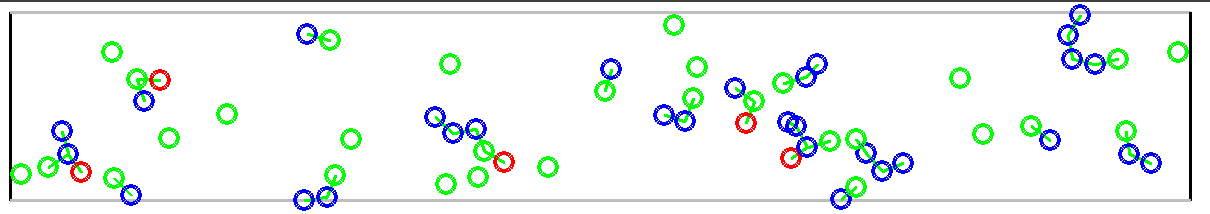}
        \caption{}
        \label{fig:efig2}
    \end{subfigure}
    \par \medskip
    \begin{subfigure}{\linewidth}
        \centering
        \includegraphics[width=.9\linewidth]{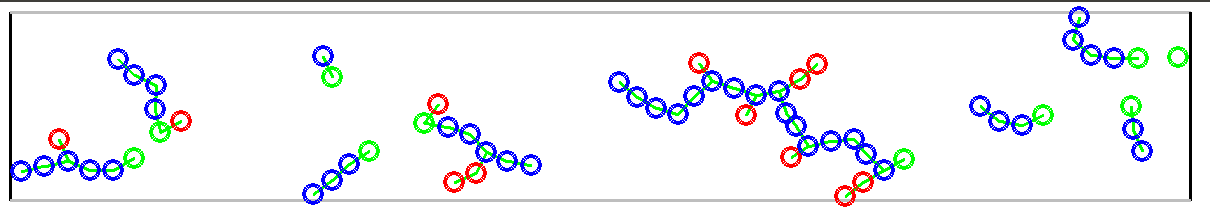}
        \caption{}
        \label{fig:efig3}
    \end{subfigure}
    \par \medskip
    \begin{subfigure}{\linewidth}
        \centering
        \includegraphics[width=.9\linewidth]{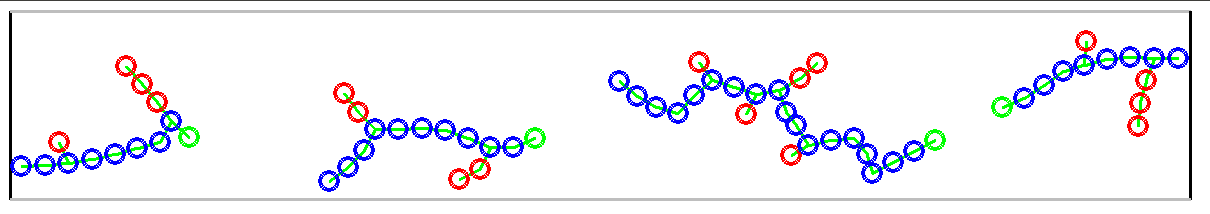}
        \caption{}
        \label{fig:efig4}
    \end{subfigure}
    \par \medskip
    \begin{subfigure}{\linewidth}
        \centering
        \includegraphics[width=.9\linewidth]{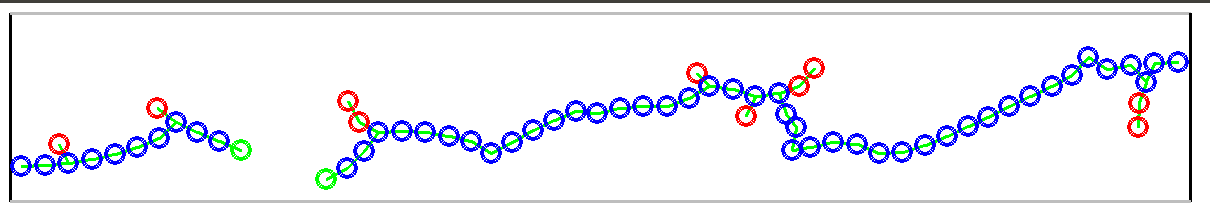}
        \caption{}
        \label{fig:efig5}
    \end{subfigure}
    \par \medskip
    \begin{subfigure}{\linewidth}
        \centering
        \includegraphics[width=.9\linewidth]{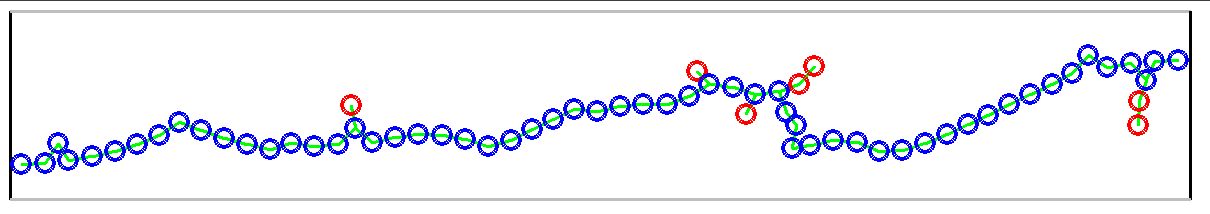}
        \caption{}
        \label{fig:efig6}
    \end{subfigure}
    \caption{Example of 1-barrier formation}
    \label{fig:formation_example}
\end{figure}

\par
 Figure \ref{fig:formation_example} graphically shows the complete algorithm running on an example scenario that is generated by simulating the algorithm. The example scenario has a belt region of length 50m and width 8m. It has 65 uniformly randomly distributed mobile sensors each having a sensing radius of 0.5m. The color coding of the sensors is the same as in Figure \ref{fig:flattening}. After the Initialization Phase, the sensor positions are as shown in Figure \ref{fig:efig1}. As seen in the figure, only small chain graphs exist spread all over the belt region, with no discernible parts of the barrier. Figure \ref{fig:efig2} shows the state after the Left-Attach Phase and Right-Attach Phase. Dominant points for all the chain graphs are calculated and are pulled towards their corresponding co-dominant points, pulling the chains with them. We can also observe some merged chain graphs compared to Figure \ref{fig:efig1}. After few steps in the simulation, we can see larger chain graphs in the belt region formed gradually as shown in Figure \ref{fig:efig3} and Figure \ref{fig:efig4}. It can be seen that there are only two chain graphs remaining (left and right chain graphs) in Figure \ref{fig:efig5} which then combine to form the final barrier as shown in Figure \ref{fig:efig6}. Note that the final barrier is non-linear and some of the sensors are still redundant.
 
 The following theorem can be proved for the correctness of the algorithm; we omit the proof here due to lack of space.
 \begin{theorem}
 \textit{For a belt region of length $L$ and sensors with sensing radius $R_s$, the proposed algorithm always terminates with a strong barrier coverage if the number of sensors is greater than or equal to $\lceil \frac{L}{2 R_s} \rceil$.}
 \end{theorem}

\subsection{Performance Evaluation}
\label{results}
 The centralized algorithm is simulated using the  \textit{pymunk}\footnote{http://pymunk.readthedocs.org/en/latest/} (python physics simulation library) and \textit{pygame}\footnote{http://www.pygame.org/} (python game development library used for visualizing sensor movements). Each sensor is modeled as a point rigid body in a gravity free environment. The connections between chain links are are modelled as \textit{slide joints}. A slide joint is like an imaginary rod between two ends, which can keep the ends from getting too far apart but will allow them to get closer to each other. The constraints for the leftmost and the rightmost sensors are modelled as \textit{Groove Joints}. The anchor point in the groove joint can slide on a specified line, which helps the leftmost and the rightmost sensors to slide on the left and right boundary when required. Pymunk allows us to apply forces on mobile rigid bodies and can simulate the results in time steps. The coordinates from pymunk simulation space are used in pygame to draw sensors and the connections between them.
\par
 The proposed algorithm is compared with two existing algorithms, the \textit{CBarrier} algorithm proposed by Shen et al. \cite{shen2008barrier}, and the \textit{CBGB} algorithm proposed by Ban et al. \cite{Ban2010}. The \textit{CBarrier} algorithm is chosen as the \textit{DBarrier} algorithm proposed in the same work is the only one that forms a truly non-linear barrier, and it has been shown that \textit{CBarrier} has a better performance than \textit{DBarrier}. The  \textit{CBGB} algorithm is chosen as it has been shown to find a near-optimal movement strategy for linear barrier formation. We simulate both algorithms and measure the average and maximum displacement of a sensor.  It should be noted that the algorithm can be centrally executed given the initial locations of the sensors and only final locations can be sent to the sensors directly without the need of sensors to make any intermediate movement. 
 
\begin{figure}[!t]
    \begin{subfigure}{\linewidth}
        \centering
        \includegraphics[width=.9\linewidth]{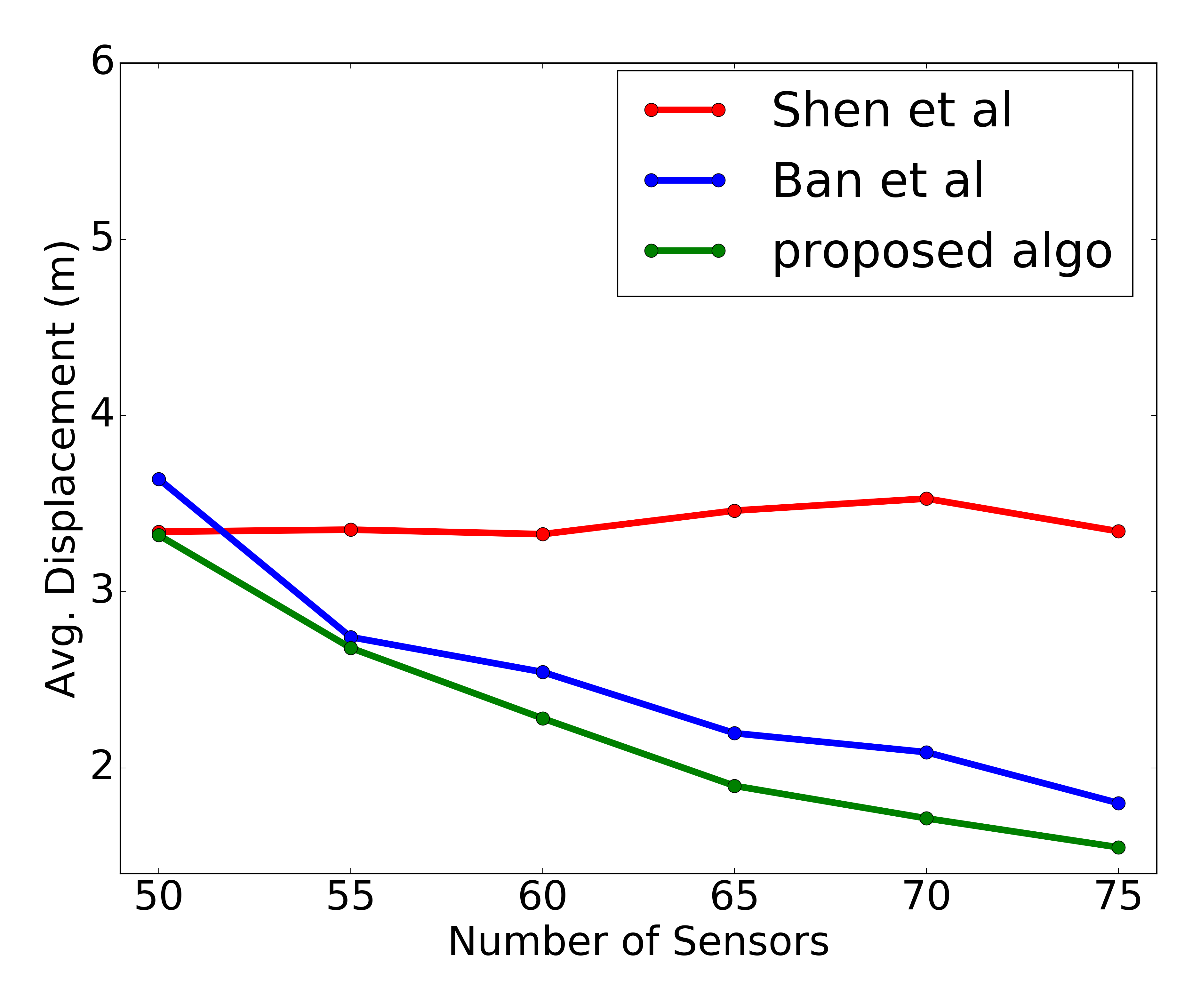}
        \caption{Average displacement}
        \label{fig:avg_cent}
    \end{subfigure}
    \par \medskip
    \begin{subfigure}{\linewidth}
        \centering
        \includegraphics[width=.9\linewidth]{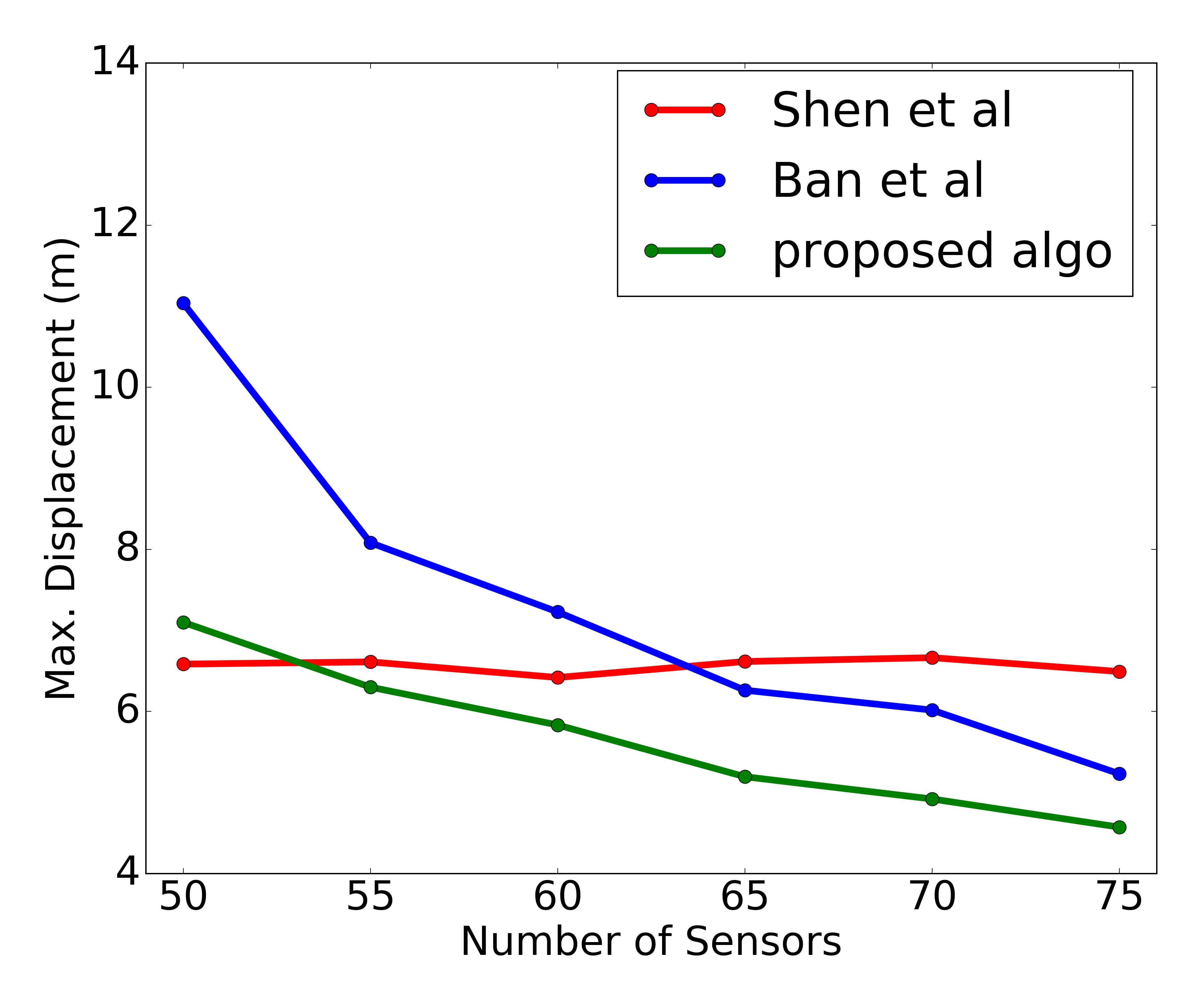}
        \caption{Maximum displacement}
        \label{fig:max_cent}
    \end{subfigure}
    \caption{Average and maximum displacement of a sensor}
    \label{fig:cent_results}
\end{figure}

 A belt of dimensions $50m \times 8m$ is taken, with varying number of sensors. The value of the sensing radius $R_s$ is taken as $0.5m$. Note that these parameter values imply that a minimum of 50 sensors are needed to provide barrier coverage (when a linear barrier is formed). The results reported are the average of running the algorithms on 100 random initial deployments. The results are shown in Figure \ref{fig:cent_results}. 
 
 The results indicate that the proposed algorithm outperforms both \textit{CBarrier} and \textit{CBGB} algorithms with respect to both the average and maximum displacement of  a sensor. In particular, the maximum displacement of the proposed algorithm is significantly better than both the algorithms. As the number of sensors are increased, the performance of the algorithm improves as more and more redundant sensors are available for local flattening and merging of chain graphs, thereby reducing the movement required.

\section{Conclusion}
\label{conclusion}
In this paper, we have presented a centralized algorithm to form a barrier from a random initial deployment using mobile sensors. The algorithm uses some novel techniques of viewing a barrier as a chain and applying laws of physics to create non-linear barriers that reduce the movement of sensors. Simulation results indicate that the algorithm outperforms other existing algorithms in terms of average and maximum displacement even with a small number of redundant sensors. The work can be extended further to investigate the design of distributed algorithms for the problem and algorithms for localized maintenance of the barrier after failures using similar techniques.


%



\ifCLASSOPTIONcaptionsoff
  \newpage
\fi



\bibliographystyle{IEEEtran}
\bibliography{citations}
%

%




\end{document}